\documentclass[review]{elsarticle}
\newcommand{\sect}[1]{\setcounter{equation}{0}\section{#1}}
\renewcommand{\theequation}{\arabic{section}.\arabic{equation}}
\newcommand{\bfm}[1]{\mbox{\boldmath${#1}$}}
\newcommand{\app}{\setcounter{section}{0}
\setcounter{equation}{0}
\renewcommand{\thesection}{APPENDIX}\renewcommand{\theequation}{\Alph{section}.\arabic{equation}}}

\begin{document}
\title{Generalized Pearson distributions for
 charged particles interacting with an electric and/or a magnetic field}
\author[ar]{A. Rossani}\ead{alberto.rossani@polito.it}
\author[ams]{A.M. Scarfone}\ead{antonio.scarfone@polito.it}
\address[ar]{Istituto Nazionale per la Fisica della Materia
(CNISM--INFM), Dipartimento di Fisica, Politecnico di
Torino, I-10129, Italy}
\address[ams]{Dipartimento di Fisica and Istituto Nazionale per la
Fisica della Materia (CNR--INFM), Sezione del Politecnico di Torino, I-10129, Italy}

\date{\today}

\begin {abstract}
The linear Boltzmann equation for elastic and/or inelastic
scattering is applied to derive the distribution function of a
spatially homogeneous system of charged particles spreading in a host medium
of two-level atoms and subjected to external electric and/or
magnetic fields. We construct a Fokker-Planck approximation to the
kinetic equations and derive the most general class of distributions
for the given problem by discussing in detail some physically meaningful
cases. The equivalence with the transport theory of electrons in a phonon
background is also discussed.
\end {abstract}


\maketitle


\sect{Introduction}

It is nowadays widely accepted that the Boltzmann equation
constitutes a very powerful mathematical model to study different
kinetic processes like fluid-dynamic problems, chemical and
nuclear reactions, diffusion and others \cite{Cercignani}.
Notwithstanding, the complicate mathematical structure embodied in
this equation makes hard to obtain explicit solutions for specific
problems, both from an analytical or a numerical point of view.\\ In
\cite{Spiga}, it has been developed a formalism to introduce
consistently the inelastic interactions in the Boltzmann equation.
In particular, it was considered a mixture of {\em test particles}
(TP) which spread in a medium of {\em field particles} (FP) endowed with two
levels of internal energy. In this way, under suitable hypotheses, a
 system describing the transport equation
for the diffusion of TP in the medium has been derived.\\
A particular simple assumption, but still preserving physical
interest, that the mass $M$ of FP is much greater than the mass $m$
of TP, gives rise to a model in which the TP can gain or
loose a fixed amount of energy. \\
In this paper, we present a generalization of the results given in
\cite{Spiga} by considering the case of charged TP diffusing in a medium
(still made by FP endowed with two energy levels), whose
interactions are modeled not only by means of an inelastic collision
integral (as in \cite{Spiga}) but also by the presence of
elastic collision.\\ In the most general fashion, we assume
the presence of an external electric and/or magnetic field
interacting with the system.\\
In the case of TP interacting with FP only by means of elastic
scattering, it is usual to adopt a Fokker-Planck approximation for
the collision integral. Such approximation leads to a solvable equation for the
distribution function (see \cite{Holt,Liboff}). As it will be shown in this paper,
the same approximation can be employed also in presence of the inelastic scattering, which leads
again to a solvable equation for the distribution function.\\ Depending on
the type of interactions occurring in the system, we obtain several physically
relevant distributions.
Among them, the generalized Pearson distribution \cite{Pearson,Johnson}
and the Margenau-Druyvesteyn distribution
\cite{Liboff} are obtained in the case of hard sphere interactions, whilst power-law distributions \cite{Buyukkilic1,Buyukkilic2}
and modified power-law distributions
\cite{Quarati} are derived in the case of Maxwellian interactions.\\
Under the above mentioned hypotheses on the masses of TP and FP, the
theory we are presenting can describe the diffusion of light
particles in an heavy medium,
so that both a loss and a gain of a fixed amount of energy is
possible. In this way, we can
show how the present model is connected with the
transport theory of electrons in a semiconductor lattice
\cite{Majorana1,Majorana2}.\\
The paper is organized as follows. Section II is devoted to the
description of the physical situation we deal with. In Section III,
by employing the truncated spherical harmonic expansion (P$_1$
approximation), we derive a system for the first two components
$N(v)$ and ${\bfm J}(v)$ of the distribution function $f({\bfm v})$.
Then, by using the same method which is commonly applied to in the presence of elastic collisions only, we construct in Section IV a Fokker-Plank approximation for the general case containing both the
elastic and the inelastic terms. Explicit distributions
corresponding to hard sphere and Maxwell interaction law, in
the presence of electric and/or magnetic fields and elastic and/or inelastic
scattering, are derived in Section V, whilst in Section VI, we obtain the expression of the particles and heat currents related to some of these distributions. In Section
VII, we study the proprieties of the  inelastic collision integral only and discuss some
aspects like the collision invariants and the trend to equilibrium.
Finally, in Section VIII, we show the mathematical equivalence of the
Boltzmann equation without the elastic collisions and the
transport theory of electrons in a phonon background. Conclusive
comments are reported in Section IX, whilst an Appendix contains
a detailed information on the distribution functions obtained in this
paper.


\sect{Outline of the problem}

Consider a spatially homogeneous medium of FP with mass $M$, endowed
with one excited internal energy level. We call $\Delta E>0$ the gap
of internal energy between the excited and the fundamental level.
Through this medium we consider TP, endowed with mass $m$ and
charge ${\cal Q}$, which diffuse in the presence of an external
electric field ${\bfm E}$ and/or magnetic field ${\bfm B}$. The TP are
supposed to interact with the medium according to the following
scheme
\begin{equation}
{\rm TP}+{\rm FP}_1\rightleftharpoons {\rm TP}+{\rm FP}_2 \ ,
\end{equation}
where FP$_1$ and FP$_2$ represent the fundamental and excited state
of FP, whose number density will be denoted by ${\cal N}_1$ and
${\cal N}_2$, respectively. The number density $n$ of TP is
considered much lower than the number density ${\cal N}={\cal
N}_1+{\cal N}_2$ of FP, so that, the medium can be
modeled as a fixed background in thermodynamical equilibrium at the
temperature $T$.\\ According to statistical mechanics we have
\begin{equation}
{{\cal N}_2\over{\cal N}_1}=\exp\left(-{\Delta E\over k\,T}\right) \ .
\end{equation}
In this case, the kinetic
equations for the distribution function of TP $f\equiv f({\bfm
x},\,{\bfm v},\,t)$ can be written as follows
\begin{equation}
{\partial\,f\over\partial\,t}+{\bfm v}\cdot{\partial\,f
\over\partial\,{\bfm x}}+{{\cal Q}\over m}\,\Big({\bf E}+{\bfm
v}\times{\bfm B}\Big)\cdot{\partial\,f\over\partial{\bfm
v}}=\left({\partial\,f\over\partial\,t}\right)^{\rm el}_{\rm
coll}+\left({\partial\,f\over\partial\,t}\right)^{\rm in}_{\rm coll}
\ .\label{Boltzmann}
\end{equation}
The elastic collision integral is given by
\begin{equation}
\left({\partial\,f\over\partial\,t}\right)^{\rm el}_{\rm
coll}=\int\int g\,I^{\rm el} (g,\,{\bfm
\Omega}\cdot{\bfm \Omega}^\prime)\Big[f({\bfm v^\prime})\, {\cal
F}({\bfm w^\prime})-f({\bfm v})\,{\cal F}({\bfm w})\Big]\,
d{\bfm w}\,d{\bfm\Omega^\prime} \ ,
\end{equation}
where
\begin{equation}
{\cal F}={\cal N}\,\left({M\over2\,\pi\,
k\,T}\right)^{3/2}\,\exp\left(-{M\,v^2\over 2\,k\,T}\right) \ ,
\end{equation}
and $I^{\rm el}(g,\,{\bfm \Omega}\cdot{\bfm \Omega}^\prime)$ is the
elastic cross section, $g=|{\bfm v}-{\bfm w}|$ is the relative
speed with $\bfm v$ and $\bfm w$ the velocities of the incoming
particles, whereas the post-collision velocities are given by
\begin{eqnarray}
&&{\bfm v^\prime}={1\over2\,(m+M)}\,\left(m\,{\bfm v}+M\,{\bfm
w}+M\,g\,{\bfm\Omega^\prime}\right) \ ,\\
&&{\bfm
w^\prime}={1\over2\,(m+M)}\,\left(m\,{\bfm v}+M\,{\bfm
w}-M\,g\,{\bfm\Omega^\prime}\right) \ .
\end{eqnarray}
Moreover, ${\bf \Omega}=({\bfm v}-{\bfm w})/g$ is the unit vector in the
direction of the relative speed, whilst the two-dimensional unit sphere $S^2$
is the domain of integration for the unit vector ${\bfm \Omega}^\prime$.\\
Differently, for the inelastic collision integral we assume that $M$ is much larger than $m$ so that the
velocity distribution function ${\cal F}_k({\bfm v})\equiv{\cal
F}_k({\bfm x},\,{\bfm v},\,t)$ for FP$_k$, at equilibrium, can be approximated by the
still-particle distribution given by
\begin{equation}
{\cal F}_k({\bfm v})={\cal N}_k\,\delta({\bfm v}) \ ,
\end{equation}
that is a Maxwellian with $M\to\infty$.\\
In this way, the inelastic collision integral becomes \cite{Spiga}
\begin{eqnarray}
\nonumber \left({\partial\,f\over\partial\,t}\right)^{\rm in}_{\rm
coll}&=&{1\over v}\int\Big[{\cal N}_1\,v_+^2\, I^{\rm
in}(v_+,\,{\bfm \Omega}\cdot{\bfm \Omega}^\prime)\,f(v_+\,
{\bfm\Omega^\prime})\\
\nonumber
& &+{\cal N}_2\,v^2\,I^{\rm in}(v,\,{\bfm \Omega}
\cdot{\bfm
\Omega}^\prime)\,f(v_-\,{\bfm\Omega^\prime})\,U(v_-^2)\Big]
\,d{\bfm\Omega^\prime}\\
\nonumber
& &-{f({\bfm v})\over v}\int\Big[{\cal N}_2\,v_+^2\,I^{\rm
in}(v_+,\,{\bfm \Omega} \cdot{\bfm \Omega^\prime})\\
& &+{\cal
N}_1\,v^2\,I^{\rm in}(v,\,{\bfm \Omega}
\cdot{\bfm\Omega^\prime})\,U(v_-^2)\Big]\,d{\bfm \Omega^\prime} \
,\label{iin}
\end{eqnarray}
where $U(x)$ is the step function, $v_\pm=\sqrt{v^2\pm\eta}$ with
$\eta=2\,\Delta E/m$ and $I^{\rm
in}(v,\,{\bfm\Omega}\cdot{\bfm\Omega^\prime})$ is the inelastic
cross section of the TP-FP$_1$ collision.
In Eq. (\ref{iin}), the first integral represents the gain of
particles with velocity $\bfm v$ due to the collisions of particles
with velocity ${\bfm v}_\pm$ and unexcited/excited scatterers,
respectively, whilst the second integral represents the loss of
particles with velocity $\bfm v$
due to collisions with unexcited/excited scatterers, respectively.\\In the following, we assume
\begin{equation}
\lim_{v\to 0}v^2\,I^\alpha(v,\,{\bfm\Omega}\cdot{\bfm\Omega^\prime})=0 \ ,
\end{equation}
with $\alpha=$ `el' or `in', which allows to perform the Fokker-Planck approximation both in presence of the elastic and the inelastic collisions. These conditions are surely fulfilled for the hard sphere and the Maxwellian interactions examined in this work.\\


\sect{Spherical harmonic expansion}

As usual \cite{Holt}, if both the spatial gradients and the electric
field are small enough, we may adopt the P$_1$ approximation, which is, basically, a weighted residual
method where the trial solution is given by a truncated spherical harmonic expansion for the distribution function of TP with shape functions 1 and ${\bfm \Omega}$:
\begin{equation}
f({\bfm v})=N(v)+{\bfm\Omega}\cdot{\bfm J}(v) \ ,
\end{equation}
where
\begin{equation}
N(v)={1\over4\,\pi}\int f(\bfm v)\,d{\bfm \Omega} \ ,\label{N}
\end{equation}
and
\begin{equation}
{\bfm J}(v)={3\over4\,\pi}\int{\bfm\Omega}\,f({\bfm
v})\,d{\bfm\Omega} \ .\label{J}
\end{equation}
Starting from the Boltzmann equation (\ref{Boltzmann}), we  derive
the evolution equations for the unknowns $N(v)$ and ${\bfm J}(v)$
by projecting it over the weights 1 and ${\bfm \Omega}$ (Galerkin method) as follows
\begin{eqnarray}
\nonumber
&&{\partial\over\partial\,t}N(v)+{v\over3}\,\nabla\cdot{\bfm
J}(v)+\tilde{\bfm
e}\cdot{1\over3\,v^2}\,{\partial\,\over\partial\,v}\Big[v^2\,\bfm{J}(v)\Big]
=\left({\partial\,N\over\partial\,t}\right)^{\rm el}_{\rm
coll}+\left({\partial\,N\over\partial\,t}\right)^{\rm in}_{\rm coll}
\ ,\\ \label{N1}\\
\nonumber
&&{\partial\over\partial\,t}{\bfm J}(v)+v\,\nabla N(v)+\tilde{\bfm
e}\,{\partial\,\over\partial\, v}\,N(v)+\tilde{\bfm b}\times{\bfm
J}(v)=\left({\partial\,{\bfm J}\over\partial\,t}\right)^{\rm
el}_{\rm coll}+\left({\partial\,{\bfm
J}\over\partial\,t}\right)^{\rm in}_{\rm coll} \ ,\\ \label{J1}
\end{eqnarray}
where we set $\tilde{\bfm e}={\cal Q}\,{\bfm E}/m$ and
$\tilde{\bfm b}={\cal Q}\,{\bfm B}/m$.\\ The expression of the
elastic collision integrals are well-known in literature and we
refer to the relevant text-books (see, for instance, Ref.
\cite{Holt}).\\ Instead, the inelastic collision integrals become
\begin{eqnarray}
\nonumber \left({\partial\,N\over\partial\,t}\right)^{\rm in}_{\rm
coll}&=&{1\over v}\Big[N(v_+)\,{\cal N}_1\,v_+^2\,I_0^{\rm
in}(v_+)+N(v_-)\,{\cal
N}_2\,v^2\,I_0^{\rm in}(v)\,U(v_-^2)\Big]\\
&-&{1\over v}\,N(v)\,\Big[{\cal N}_2\,v_+^2\,I_0^{\rm in}(v_+)
+{\cal N}_1\,v^2\,I_0^{\rm in}(v)\,U(v_-^2)\Big] \ ,\\
\nonumber \left({\partial\,{\bfm J}\over\partial\,t}\right)^{\rm
in}_{\rm coll}&=&{1\over v}\Big[{\bfm J}(v_+)\,{\cal
N}_1\,v_+^2\,I_1^{\rm in}(v_+)+{\bfm J}(v_-)\,{\cal
N}_2\,v^2\,I_1^{\rm
in}(v)\,U(v_-^2)\Big]\\
&-&{1\over v}\,{\bfm J}(v)\,\Big[{\cal N}_2\,v_+^2\,I_0^{\rm
in}(v_+)+{\cal N}_1\,v^2\,I_0^{\rm in}(v)\,U(v_-^2)\Big] \ ,
\end{eqnarray}
where
\begin{equation}
I_\ell^{\rm in}(v)=2\,\pi\int\limits_{-1}\limits^1 I^{\rm
in}(v,\,\mu)\,\mu^\ell\,d\mu \ ,
\end{equation}
with $\ell=0,\,1$ and $\mu={\bfm\Omega}\cdot{\bfm\Omega^\prime}$.\\
In view of the manipulations we are going to perform, by assuming a
stationary and space independent case, it is convenient to introduce the auxiliary quantities
$\xi=v^2$, $F(\xi)=N(v)$ and ${\bfm G}(\xi)={\bfm J}(v)$, so that
Eqs. (\ref{N1}) and (\ref{J1}) become
\begin{eqnarray}
&&\tilde{\bfm
e}\cdot{2\over3\sqrt{\xi}}\,{\partial\over\partial\,\xi}\Big[\xi\,{\bfm
G}(\xi)\Big] =\left({\partial\,F\over\partial\,t}\right)^{\rm
el}_{\rm coll}+\left({\partial\,F\over\partial\,t}\right)^{\rm
in}_{\rm
coll} \ ,\\
&&\tilde{\bfm
e}\,2\,\sqrt{\xi}\,{\partial\over\partial\,\xi}\,F(\xi)+\tilde{\bfm
b}\times{\bfm G}(\xi)=\left({\partial\,{\bfm
G}\over\partial\,t}\right)^{\rm el}_{\rm
coll}+\left({\partial\,{\bfm G}\over\partial\,t}\right)^{\rm
in}_{\rm coll} \ ,
\end{eqnarray}
where
\begin{eqnarray}
\nonumber \left({\partial\,F\over\partial\,t}\right)^{\rm in}_{\rm
coll}&=& {1\over\sqrt{\xi}}\Big[F(\xi+\eta)\,{\cal
N}_1\,(\xi+\eta)\,\sigma_0^{\rm in}(\xi+\eta)\\
\nonumber
& &+F(\xi-\eta)\,{\cal
N}_2\,\xi\,\sigma_0^{\rm in}(\xi)\,U(\xi-\eta)\Big]\\
\nonumber
& &-{F(\xi)\over\sqrt\xi}\,\Big[{\cal
N}_2\,(\xi+\eta)\,\sigma_0^{\rm in}(\xi+\eta)+{\cal
N}_1\,\xi\,\sigma_0^{\rm in}(\xi)\,U(\xi-\eta)\Big] \ ,\\
\label{F}\\
\nonumber\left({\partial\,{\bfm G}\over\partial\,t}\right)^{\rm
in}_{\rm coll}&=&{1\over\sqrt\xi}\Big[{\bf G}(\xi+\eta)\,{\cal
N}_1\,(\xi+\eta)\,\sigma_1^{\rm
in}(\xi+\eta)\\
\nonumber
& &+{\bfm G}(\xi-\eta)\,{\cal N}_2\,\xi\,\sigma_1^{\rm in}(\xi)\,U(\xi-\eta)\Big]\\
\nonumber
& &-{{\bfm G}\over\sqrt{\xi}}\,\Big[{\cal
N}_2\,(\xi+\eta)\,\sigma_0^{\rm in}(\xi+\eta)+{\cal
N}_1\,\xi\,\sigma_0^{\rm in}(\xi)\,U(\xi-\eta)\Big] \ ,\\
\label{GG}
\end{eqnarray}
and $\sigma_\ell^{\rm in}(\xi)=I_\ell^{\rm in}(v)$.


\sect{The Fokker-Planck collision term}

In the following we extend a standard procedure, often employed to derive the
Fokker-Planck approximation of $\left(\partial\,f/\partial\,t\right)^{\rm el}_{\rm
coll}$, to the case of inelastic collisions.\\
We assume
\begin{equation}
\left({k\,T\over M}\right)^2<<\eta^2\sim {k\,T\over M}\,{k\,T\over
m}\,<<\left({k\,T\over m}\right)^2 \ ,
\end{equation}
which means that $\eta$ is considered small with respect to the
thermal square speed of a TP but still large with respect to the
thermal square speed of a FP.

Under this
assumption we can adopt a procedure which leads to the Fokker-Planck
approximation of $(\partial\,F/\partial\,t)^{\rm in}_{\rm coll}$. In
fact, by considering an arbitrary smooth function $\Phi(\xi)$, one
can easily show that
\begin{eqnarray}
\nonumber &
&\int\limits_0\limits^\infty\sqrt\xi\,\left({\partial\,F\over\partial\,t}\right)^{\rm
in}_{\rm
coll}\,\Phi(\xi)\,d\xi\\
&=&\int\limits_0\limits^\infty
(\xi+\eta)\,\sigma_0^i(\xi+\eta)\,\Big[{\cal N}_1\,F(\xi+\eta)-{\cal
N}_2\,F(\xi)\Big]\,\Big[\Phi(\xi)-\Phi(\xi+\eta)\Big]\,d\xi \ .
\end{eqnarray}
We expand the integrand at the right hand side in power series
of $\eta$ and retain only the terms up to $\eta^2$, so that
\begin{eqnarray}
\nonumber
& &\int\limits_0\limits^\infty\sqrt\xi\,\left({\partial\,F\over\partial\,t}\right)^{\rm
in}_{\rm coll}\,\Phi(\xi)\,d\xi\\
\nonumber
&=&-{\eta^2\over2}\,{\cal
N}\int\limits_0\limits^\infty \xi\, \sigma_0^i(\xi)\,
\Bigg[{m\over2\,k\,T}\,F(\xi)+{\partial\over\partial\,\xi}\,F(\xi)\Bigg]
\,{\partial\over\partial\,\xi}\,\Phi(\xi)\,d\xi\\
\nonumber &=&{\eta^2\over2}\,{\cal N}\int\limits_0\limits^\infty
{\partial\
\over\partial\xi}\left\{\xi\,\sigma_0^i(\xi)\,\Bigg[{m\over2\,k\,T}\,F(\xi)
+{\partial\over\partial\,\xi}\,F(\xi)\Bigg]\right\}\,\Phi(\xi)\,d\xi
\ ,\\
\end{eqnarray}
where we integrated by parts. Since $\Phi(\xi)$ is arbitrary we
can set
\begin{equation}
\left({\partial\,F\over\partial\,t}\right)^{\rm in}_{\rm
coll}={\eta^2\over2}\,{{\cal N}\over\sqrt\xi}\,{\partial\
\over\partial\,\xi}\left\{\xi\,\sigma_0^i(\xi)\,\Bigg[{m\over2\,k\,T}\,F(\xi)
+{\partial\over\partial\,\xi}\,F(\xi)\Bigg]\right\} \ ,\label{jin}
\end{equation}
and from the stationary condition $(\partial\,F/\partial\,t)^{\rm
in}_{\rm coll}=0$ we derive the following equilibrium distribution function for
the present approximation
\begin{equation}
F(\xi)=C\,\exp\left(-{m\,\xi\over2\,k\,T}\right) \ ,
\end{equation}
that is a Maxwellian with $C$ the normalization constant.\\
Let us now observe that Eq. (\ref{GG}), together with Eq.
(\ref{jin}), shows that ${\bfm G}(\xi)={\cal O}(\eta^2)$ so that we
can approximate $(\partial\,{\bfm G}/\partial\,t)^{\rm in}_{\rm
coll}$ in
\begin{equation}
\left({\partial\,{\bfm G}\over\partial\,t}\right)^{\rm in}_{\rm
coll}=-{\cal N}\,\sqrt\xi\,{\bfm G}(\xi)\,\sigma^{\rm in}(\xi) \
,\label{bjin}
\end{equation}
where
\begin{equation}
\sigma^{\rm in}(\xi)=\sigma_0^{\rm in}(\xi)-\sigma_1^{\rm in}(\xi) \
,
\end{equation}
is the transport cross section for the inelastic process.\\ Now, we want
now utilize the approximate interaction terms (\ref{jin})
and (\ref{bjin}) when both inelastic and elastic collisions occur.\\
It is known \cite{Holt} that for $m<<M$ the Fokker-Planck elastic
collision terms reads
\begin{equation}
\left({\partial\,F\over\partial\,t}\right)^{\rm el}_{\rm
coll}={2\over M}\,{{\cal N}\over\sqrt\xi}\,{\partial
\over\partial\,\xi}\left\{\xi^2\,\left[
m\,F(\xi)+2\,k\,T\,{\partial\over\partial\,\xi}\,F(\xi)\right]\,\sigma_0^{\rm
el}(\xi)\right\} \ ,
\end{equation}
and
\begin{equation}
\left({\partial\,{\bfm G}\over\partial\,t}\right)^{\rm el}_{\rm
coll}=-{\cal N}\,\sqrt\xi\,{\bfm G}(\xi)\,\sigma^{\rm el}(\xi) \ ,
\end{equation}
where
\begin{equation}
\sigma^{\rm el}(\xi)=\sigma_0^{\rm el}(\xi)-\sigma_1^{\rm el}(\xi) \
,
\end{equation}
is the transport cross section for the elastic process.\\ In
particular, if $\sigma^{\rm in}(\xi)\sim\sigma^{\rm el}(\xi)$, it is
actually possible to add together the two collision terms. In fact,
we can say that the order of approximation is the same if
$\eta^2\sim k\,T\,\xi/M\sim (k\,T)^2/(M\,m)$, which corresponds
exactly to our assumption on $\eta^2$.\\ Finally, by collecting the
effects of elastic and inelastic interactions we obtain the
following equations
\begin{eqnarray}
\nonumber
{2\over3}\,{\bfm e}\cdot{\partial\over\partial\,\xi}\Big[\xi\,{\bfm
G}(\xi)\Big]&=&{\partial\over\partial\,\xi}
\left\{\left[{\eta^2\over2}\,\,\xi\,\sigma_0^{\rm
in}(\xi)+{4\,k\,T\over M}\,\xi^2\,\sigma^{\rm el}(\xi)\right]\right.\\
& &\times\left.
\left[{m\over2\,k\,T}\,F(\xi)+{\partial\over\partial\,\xi}\,F(\xi)\right]\right\}
\ ,\label{prima}
\end{eqnarray}
and
\begin{equation}
2\,{\bfm e}\,{\partial\over\partial\,\xi}\,F(\xi)+{{\bfm
b}\over\sqrt\xi}\times{\bfm G}(\xi)=-{\bfm G}(\xi)\,\sigma(\xi) \,
\end{equation}
with $\sigma(\xi)=\sigma^{\rm in}(\xi)+\sigma^{\rm el}(\xi)$, ${\bfm
e}=\tilde{\bfm e}/{\cal N}$ and ${\bfm b}=\tilde{\bfm b}/{\cal
N}$.\\
We can solve this last equation for ${\bfm G}(\xi)$ to obtain
\begin{eqnarray}
{\bfm
G}(\xi)=-2\,\left[\sigma^2(\xi)+{b^2\over\xi}\right]^{-1}{\partial\over\partial\,\xi}\,F(\xi)
\,\left({{\bfm e}\times{\bfm b}\over\sqrt\xi}+{{\bfm e}\cdot{\bfm
b}\over\xi\,\sigma(\xi)}\,{\bfm b}+\sigma(\xi)\,{\bfm e}\right) \ ,\label{g}
\end{eqnarray}
that, introduced in Eq. (\ref{prima}), gives
\begin{eqnarray}
\nonumber & &
\left\{{4\over3}\,\left[\sigma^2(\xi)+{b^2\over\xi}\right]^{-1}\left[{({\bfm
e}\cdot{\bfm
b})^2\over\xi\,\sigma(\xi)}+\sigma(\xi)\,e^2\right]\right.\\
&+&\left.{4\,k\,T\over
M}\,\xi^2\,\sigma^{\rm el}(\xi)+{\eta^2\over2}\,\,\xi\,\sigma_0^{\rm
in}(\xi)\right\}\,{\partial\over\partial\,\xi}\,F(\xi)\\
=&-&{m\over2\,k\,T}\left[{4\,k\,T\over M}\,\xi\,\sigma^{\rm
el}(\xi)+{\eta^2\over2}\,\sigma_0^{\rm in}(\xi)\right]\,F(\xi) \ .
\end{eqnarray}
This equation is easily integrated, giving
\begin{equation}
F(\xi)=C\,\exp\left(-{m\over2\,k\,T}\,\Xi(\xi)\right) \
,\label{distr}
\end{equation}
where the constant $C$ is obtained by the normalization condition
\begin{equation}
2\,\pi\int\limits_0\limits^\infty\sqrt\xi\,F(\xi)\,d\xi=n \ ,
\end{equation}
and the function $\Xi(\xi)$ is given by
\begin{equation}
\Xi(\xi)=\int\limits_0\limits^\xi\left[{4\,k\,T\over M}\,x\,\sigma^{\rm
el}(x)+{\eta^2\over2}\,\sigma_0^{\rm
in}(x)\right]\Big/ {\cal F}(x)\,dx \ ,\label{xi}
\end{equation}
with
\begin{equation}
F(x)={4\over3}\,\Big(x\sigma^2(x)+{b^2
}\Big)^{-1}\Big[{\big({\bfm e}\cdot{\bfm b}\big)^2\over\sigma(x)}
+e^2x\,\sigma(x)\Big]+{4\,k\,T\over M}\,x\,\sigma^{\rm
el}(x)+{\eta^2\over2}\,\sigma_0^{\rm in}(x) \ .
\end{equation}
The distribution (\ref{distr}) contains a rich variety of cases, some
of them known in the literature, like, for instance, the Margenau-Druyvesteyn distribution \cite{Holt,Liboff} the power-law distribution \cite{Buyukkilic1,Buyukkilic2}, the modified power-law distribution \cite{Quarati} and, for $e=0$, the Maxwell distribution.

\sect{Some physically meaningful cases}

Under certain assumptions on the cross sections we can obtain the
explicit expression of the distribution function (\ref{distr}). In particular, we
consider two physically meaningful interaction
laws which give rise to different situations.\\
The first case is given by the hard sphere interactions, by
assuming constant values for all the cross sections, with
$\sigma^{\rm el}(\xi)\equiv\sigma^{\rm el}$ and $\sigma^{\rm
in}(\xi)\equiv\sigma^{\rm in}$. Equation (\ref{xi}) becomes
\begin{equation}
\Xi(\xi)=\int\limits_0\limits^\xi\left[1+{a_1\,x+a_2\over(a_3\,x+a_4)
\,(a_5\,x+a_6)}\right]^{-1}\,dx \ ,
\end{equation}
with $a_1=8\,\sigma\,e^2/3$, $a_2=8\,({\bfm e}\cdot{\bfm b})^2/(3
\,\sigma)$, $a_3=8\,k\,T\,\sigma^{\rm el}/M$,
$a_4=\eta^2\,\sigma_0^{\rm in}$, $a_5=\sigma^2$ and $a_6=b^2$.\\
This integral can be easily calculated, obtaining the distribution
\begin{equation}
F_1(\xi)=C_1\,\Big(1+b_1\,\xi+b_2\,\xi^2\Big)^{1/b_3}\,\exp\Big(-b_4\,\xi-b_5\,
\arctan(b_6+b_7\,\xi)\Big) \ ,\label{distr1}
\end{equation}
which depends on the
parameters $b_1,\,\ldots,\,b_7$, whose relationships with the
microscopic quantities of the system are given in Appendix.\\
Function (\ref{distr1}) is a generalized Pearson distribution, since it
can be derived as a solution of the following linear differential equation
\begin{equation}
{d\,p(\xi)\over d\,\xi}+{w_6\,\xi^2+w_5\,\xi+w_4\over w_3\,\xi^2+w_2\,\xi+w_1}\,p(\xi)=0 \ ,
\end{equation}
for suitable values of parameters $w_1,\,\ldots,\,w_6$. The case $w_6=0$ corresponds to the original problem
defining the family of the Pearson distributions
\cite{Pearson,Johnson}.\\
Several interesting distributions can be obtained, as special cases,
from the family (\ref{distr1}). In particular, without the
elastic collisions ($\sigma^{\rm el}=0$ and
$\sigma\equiv\sigma^{\rm in}$) the distribution reduces to
\begin{equation}
F_2(\xi)=C_2\,\Big(1+c_1\,\xi\Big)^{-c_2}\,\exp\left(-c_3\,\xi\right)
\ ,\label{distr3}
\end{equation}
where the coefficients $c_i>0$ (see Appendix)
include the effects of the inelastic scattering only. This function
is a modified power-law distribution. It has been previously considered in
\cite{Quarati} from a kinetic approach, based on a nonlinear
Fokker-Planck equation taking into account
the effects of a generalized exclusion-inclusion principle \cite{Kaniadakis}.\\
Another interesting case arises from Eq. (\ref{distr1}) without the magnetic field
\begin{equation}
F_3(\xi)=C_3\,\Big(1+d_1\,\xi\Big)^{d_2}\,\exp\left(-d_3\,\xi\right) \
,\label{distr4}
\end{equation}
which is the Margenau-Druyvesteyn distribution with positive coefficients $d_i>0$,
whose expression in terms of microscopical
quantities are given in Appendix.\\
We summarize in  Table 1 the main distributions obtained from
the hard sphere interaction law in the presence of elastic and/or inelastic
collisions and in the presence of electric and/or magnetic fields. \vspace{5mm}
\begin{center}
Table 1. Classification of the distribution functions for the hard
sphere interaction case with elastic and/or inelastic collisions and electric and/or magnetic external fields.\\
\vspace{5mm}
\begin{tabular}{c||ccc}\hline
 & $\sigma^{\rm el}$ and $\sigma^{\rm in}$ & $\sigma^{\rm el}$ & $\sigma^{\rm in}$\\
\hline\hline \ \ \ ${\bfm e}$ and ${\bfm b}$ \ \ \ & \ \ Generalized-Pearson \ \ &
\ \ Generalized-Pearson \ \ & \ \
Modified Tsallis \ \ \\
\ \ \ ${\bfm e}$ \ \ \ & Druyvesteyn & Druyvesteyn & \ \
Heated Maxwellian \\
\ \ \ ${\bfm b}$ \ \ \ & Maxwellian & Maxwellian
& Maxwellian \\
\hline
\end{tabular}
\end{center}
\vspace{5mm}

The second case we study is given by the Maxwellian
interactions where cross sections are given by $\sigma^{\rm
el}(\xi)=\tilde\sigma^{\rm el}/\sqrt{\xi}$ and $\sigma^{\rm
in}(\xi)=\tilde\sigma^{\rm in}/\sqrt{\xi}$. Formula (\ref{xi}) now
becomes
\begin{equation}
\Xi(\xi)=\int\limits_0\limits^\xi\left(1+{f_1\,x\over f_2\,x+f_3
}\right)^{-1}\,dx \ ,
\end{equation}
with $f_1=8\,[({\bfm e}\cdot{\bfm
b})^2/\tilde\sigma+\tilde\sigma\,e^2]/[3\,(\tilde\sigma^2+b^2)]$,
$f_2=8\,k\,T\,\tilde\sigma^{\rm el}/M$ and
$f_3=\eta^2\,\tilde\sigma_0^{\rm in}$.\\
By performing the integration we obtain
\begin{equation}
F_5(\xi)=C_5\,\Big(1+g_1\,\xi\Big)^{-g_2}\,\exp\left(-g_3\,\xi\right) \
,\label{distr2}
\end{equation}
which is again a modified power-law distribution, depending on three positive coefficients $g_i>0$
(see Appendix). This distribution is functionally equivalent to the
distribution (\ref{distr3}), i.e. hard sphere TP with purely
inelastically interacting behaviors like Maxwellian TP.\\
An interesting situation, already derived in \cite{Rossani}, is
obtained when the scattering is purely inelastic, corresponding to
$g_3\to0$ limit. This gives rise to a power-law distribution that can be written in the form
\begin{equation}
F_6(\xi)=C_6\,\left[1-(1-q)\,{m\,\xi\over2\,k\,T}\right]^{1/(1-q)} \ ,\label{tsa}
\end{equation}
known in literature as Tsallis distribution \cite{Tsallis}. As show in Appendix,
the deformed parameter $q$ is related to the microscopic
quantities of the system according to
\begin{equation}
q=1+{16\over3}\,{k\,T\over m}\,{({\bfm e}\cdot{\bfm b})^2+e^2\,
\tilde\sigma^2\over (b^2+\tilde\sigma^2)\,\eta^2\,\tilde\sigma\,\tilde\sigma^{\rm in}_0} \ ,\label{q}
\end{equation}
which is an increasing function of the electric field and a decreasing function of the magnetic field.
(Remark that in Ref. \cite{Rossani} a different definition for
the deformed parameter $q$ has been adopted, which is related to the present one through $q\to2-q$).\\
In Table 2 we summarize the main distributions obtained from the
Maxwell interaction law in presence of elastic and/or inelastic
collisions and in presence of electric and/or magnetic fields.
\begin{center}
Table 2. Classification of the distribution functions in the Maxwell
interaction case with elastic and/or inelastic collisions and electric and/or magnetic external fields.\\
\vspace{5mm}
\begin{tabular}{c||ccc}\hline
 & $\tilde\sigma^{\rm el}$ and $\tilde\sigma^{\rm in}$ & $\tilde\sigma^{\rm el}$ & $\tilde \sigma^{\rm in}$\\
\hline\hline \ \ \ ${\bfm e}$ and ${\bfm b}$ \ \ \ & \ \  Modified Tsallis \ \
& \ \ Heated Maxwellian \ \ & \ \
Tsallis \ \ \\
\ \ \ ${\bfm e}$ \ \ \ & \ \  Modified Tsallis \ \ & \ \ Heated Maxwellian \ \ & \ \
Tsallis \ \ \\
\ \ \ ${\bfm b}$ \ \ \ & Maxwellian & \ \ Maxwellian
& Maxwellian \\
\hline
\end{tabular}
\end{center}


\sect{Physical quantities}

From a physical point of view, we are interested to study the conditions assuring the
existence of the first $\ell$ momenta of the distribution. They are defined in
\begin{equation}
\langle{\bfm v}^\ell\rangle={\int{\bfm v}^\ell\,f({\bfm v})
\,d{\bfm v}\over\int f({\bfm v})\,d{\bfm v}} \ ,\label{i}
\end{equation}
so that, for $\ell=1$, we obtain the definition of the density current, given by
\begin{equation}
{\bfm j}=n\,{\bfm u}={2\,\pi\over3}\int\limits_0\limits^\infty\xi\,{\bfm G}(\xi)\,d\xi \ ,\label{i2}
\end{equation}
with ${\bfm u}$ the mean velocity of TP while, for $\ell=2$ we have
\begin{equation}
\langle v^2\rangle={{\cal L}_3\over{\cal L}_1} \ ,\label{i3}
\end{equation}
which corresponds to the mean-squares speed,
where, for sake of convenience, we introduced the quantity
\begin{equation}
{\cal L}_\ell=\int\limits_0\limits^\infty\xi^{\ell/2}\,\phi(\xi)\,d\xi \ ,
\end{equation}
with
\begin{equation}
\phi(\xi)=\exp\left(-{m\over k\,T}\,\Xi(\xi)\right) \ .
\end{equation}
The existence of these quantities is assured for
almost all the distributions derived in the previous section, since the
presence of the exponential factor guarantees a fast convergence of the integrals.\\
An exception is given by the distribution (\ref{tsa}), which has a purely power-law asymptotic behavior.
In this case, the convergence of the above integrals, and in particular the normalization
condition, imposes a limitation on the value that the parameter $q$ can assumes.
We recall, (see \cite{Markowich} and reference therein), that in the
stationary and homogenous case the transport equation for charged particles subjected to
an external electric field has not, in general, a normalizable solution. The
nonexistence of such a solution is known as {\em runaway phenomenon} and
its occurrence depends on how fast is the decay of the cross section as
a function of the energy.\\
For the distribution (\ref{tsa}),  the convergence of the quantity $\langle v^2\rangle$ (which also guarantees the convergence of the lower momenta with $\ell<2$) is assured by the further condition
\begin{equation}
1\leq q\leq{7\over5} \ ,
\end{equation}
which defines, through Eq. (\ref{q}), a boundary in $\bfm e$ and $\bfm b$ for the onset of the runaway phenomenon.\\
Another physically relevant quantity is  the heat flux, defined by
\begin{eqnarray}
\nonumber
{\bfm q}&=&{1\over2}\,m\int({\bfm v}-{\bfm u})\,({\bfm v}-{\bfm u})^2\,f({\bfm v})\,d{\bfm v}\\
&=&n\,m\left({\cal G}-{5\over6}\,{{\cal L}_3\over{\cal L}_1}+u^2\right)\,{\bfm u} \ ,\label{qq}
\end{eqnarray}
where we have set
\begin{equation}
{\cal G}={\pi\over3\,n\,u}\int\limits_0\limits^\infty\xi^2\,G(\xi)\,d\xi \ .
\end{equation}
In general, both the quantities $\bfm j$ and $\bfm{q}$
cannot be obtained in a closed analytical form.\\
Notwithstanding, exact expressions for these currents can be obtained for the case of charged TP governed by the only inelastic scattering. A simpler assumption, still preserving
the physical interest, is given by setting $b=0$. In this case, our problem shrunk in just one dimension, since now both the currents $\bfm j$ and ${\bfm q}$ assume the same direction of the electric field. In this simple situation, for particles undergoing to hard sphere interactions, we obtain the expressions
\begin{eqnarray}
{\bfm j}&=&{4\over3}\,n\,\left({m\over2\,\pi\,k\,T_\ast}\right)^{1/2}
\,{{\bfm e}\over\sigma} \ ,\label{j1}\\
{\bfm q}&=&{n\,m\over3\,\sqrt\pi}\,\left({2\,k\,T_\ast\over m}\right)^{1/2}\,\left[
{4\over\pi}\,\left({2\over3}\,{m\over k\,T_\ast}\,{e\over\sigma}\right)^2-1\right]{{\bfm e}\over\sigma} \ ,\label{j2}
\end{eqnarray}
where $T_\ast$, derived in Appendix, is given by
\begin{equation}
T_\ast=T\,\left(1+{8\,e^2\over3\,\eta^2\,\sigma\,\sigma_0^{\rm in}}\right) \ .
\end{equation}
Let us observe that in the $e\to0$ limit the particle current shows an Ohmic behavior with ${\bfm j}\propto{\bfm e}$, while for $e\to\infty$ it approaches an asymptotic value
\begin{equation}
j_\infty=n\,\eta\left({m\,\sigma_0^{\rm in}\over3\,\pi\,k\,T\,\sigma}\right)^{1/2} \ .
\end{equation}
In the same way, in the case of Maxwell interactions, we have
\begin{equation}
{\bfm j}=n\,{{\bfm e}\over\tilde\sigma} \ ,\quad\quad{\rm and}\quad\quad
{\bfm q}=n\,m\,{e^2\over\tilde\sigma^3}\,{\bfm e} \ ,
\end{equation}
and the particle current shows always an Ohmic behavior.\\
More in general, it is known that in the limit of weak electric field, particle current and heat current can be expressed by means of the following constitutive equations
\begin{eqnarray}
{\bfm j}&=&k_{11}\,{\bfm E}+k_{12}\,{\bfm\nabla}T+k_{13}\,{\bfm\nabla}n \ ,\\
{\bfm q}&=&k_{21}\,{\bfm E}+k_{22}\,{\bfm\nabla}T+k_{23}\,{\bfm\nabla}n \ ,
\end{eqnarray}
where, the Einstein law \cite{Holt}
\begin{equation}
{k_{13}\over k_{11}}={k_{23}\over k_{21}}=-{k\,T\over n\,|{\cal Q}|} \ ,\label{ein}
\end{equation}
can be rewritten in the compact form
\begin{equation}
{\bfm j}=k_{11}\,\bfm{\tilde E}+k_{12}\,{\bfm\nabla}T \ ,\quad\quad{\rm and}\quad\quad
{\bfm q}=k_{21}\,\bfm{\tilde E}+k_{22}\,{\bfm\nabla}T \ ,
\end{equation}
where $\bfm{\tilde E}={\bfm E}-(k\,T/n\,|{\cal Q}|){\bfm\nabla}\,n$.\\
Further, we can extract informations about all the coefficients $k_{ij}$ from the Eqs. (\ref{i2}) and (\ref{qq}) although these relations hold only for the spatially homogeneous case.\\
In fact, as known, the coefficients $k_{ij}$ are related to each other by means of the following relations: \begin{equation}
{k_{21}\over k_{12}}=-T \ ,
\end{equation}
that is the Onsager law and \cite{Onsager}
\begin{equation}
{k_{22}\over k_{11}}=-{\pi^2\over3}\,\left({k\over |{\cal Q}|}\right)^2\,T \ ,
\end{equation}
that is the Weidemann-Franz law \cite{Liboff}.\\
Starting from the expression of $k_{11}$ and $k_{21}$ derived from the linearization of ${\bfm j}$ and ${\bfm q}$, we can obtain all the other coefficients.
For instance, we can linearize the expression of the currents (\ref{j1}) and (\ref{j2}), which hold for the hard sphere interactions, to obtain the coefficients
\begin{equation}
k_{11}={4\over3}\,{n\,{\cal Q}\over\sigma\,{\cal N}}\,\left(2\,\pi\,m\,k\,T\right)^{-1/2} \ ,\quad\quad{\rm and}\quad\quad
k_{21}=-{1\over3}\,{n\,{\cal Q}\over\sigma\,{\cal N}}\,\left(2\,k\,T\over\pi\,m\right)^{1/2} \ ,
\end{equation}
from which one can deduce the expressions of the remaining coefficients $k_{ij}$.\\


\sect{Study of the inelastic collision integral}

In this section, we deal with the properties and the
consequences that the inelastic collision integral induces into
the system. For this purpose, we start by considering the weak form
of the transport equation, without the elastic collisions. This can be accomplished by
considering a smooth function $\phi({\bfm v})$ and by introducing the
following auxiliary functional
\begin{eqnarray}
\nonumber
{\cal G}[\phi]&=&\int\phi(\bfm v)\,\left({\partial\,f\over\partial\,t}\right)^{\rm in}_{\rm
coll}\,d{\bfm v}\\
\nonumber &=&\int d{\bfm\Omega}\int d{\bfm\Omega^\prime}\int
\Big[\phi(v\,{\bfm\Omega}) -\phi(v_+\,{\bfm\Omega^\prime})\Big]
\Big[{\cal N}_1\,f(v_+\,{\bfm\Omega^\prime})-{\cal N}_2\,f(v\,{\bfm\Omega})\Big]\\
& &\times v_+^2\,I^{\rm in}(v_+,\,{\bfm\Omega}\cdot{\bfm\Omega^\prime})\,v\,dv \ .
\end{eqnarray}
From this equation, we find that any arbitrary function
$\phi({\bfm v})=\Phi(v^2)$ (depending only on the modulo of $\bfm v$) and
periodic with period $\epsilon^2$
\begin{equation}
\Phi(v^2+\epsilon^2)=\Phi(v^2) \ ,\label{const}
\end{equation}
implies ${\cal G}[\phi]=0$, i.e. it is a collisional invariant for the given problem.\\
In particular, the choice $\Phi=1$, a constant, corresponds to the TP number
conservation. We observe that the initial assumption
$M\to\infty$ has the important consequence that TP does not conserve
the ordinary physical quantities like energy and momentum.
Moreover, Eq. (\ref{const}) implies the existence of infinite
constants of motion (spurious invariants), where $\Phi=1$ is just the simplest case.\\
Let us now consider the above weak equation applied to the function
\begin{equation}
\phi({\bfm v})=\ln\Bigg[f({\bfm v})\exp\left({mv^2\over
kT}\right)\Bigg] \ .
\end{equation}
In this case, the auxiliary function ${\cal G}[\phi]$ becomes
\begin{eqnarray}
\nonumber
{\cal G}[\phi]&=&\int d{\bfm\Omega}\int d{\bfm\Omega^\prime}\int \Big\{\ln f(v\,{\bfm\Omega})-\ln\big[{\cal E}\,f(v_+\,{\bfm\Omega^\prime})\big]\Big\}\\
& &\times\Big[{\cal E}\,f(v_+\,{\bfm\Omega^\prime})-f(v\,{\bfm\Omega})\Big]\,{\cal
N}_2\,v_+^2\, I^{\rm
in}(v_+,\,{\bfm\Omega}\cdot{\bfm\Omega^\prime})\,v\,dv\le0 \
.\label{h}
\end{eqnarray}
Equation (\ref{h})
is a non positive quantity, as it follows from the inequality $(a-b)\,\ln(a/b)\ge 0$ and consequently it states an $H$ theorem for the present
problem.\\ At equilibrium, it must be ${\cal G}[\phi]=0$, for any $v$, $\bfm\Omega$ and
$\bfm\Omega^\prime$. This implies the
following relation
\begin{equation}
\ln f(v\,{\bfm\Omega})-\ln\big[{\cal
E}\,f(v_+\,{\bfm\Omega^\prime})\big]=0 \ ,
\end{equation}
from which one easily obtains a family of equilibrium distributions
\begin{equation}
f(v)=\Gamma(v^2)\,\exp\left(-{m\,v^2\over 2\,k\,T}\right) \ ,
\end{equation}
where $\Gamma(v^2)$ is an arbitrary function that fulfils the periodicity
condition
\begin{equation}
\Gamma(v^2+\epsilon^2)=\Gamma(v^2) \ ,
\end{equation}
originated from the condition (\ref{const}).\\
Clearly, the family of distributions (\ref{distr}) contains, as a
special case, for $\Gamma(v^2)$ a constant, the Maxwellian distribution,
whilst, in all the other cases, it generates a family of infinite
distorted Maxwellians, one for each choice of the function $\Gamma(v^2)$.


\sect{Equivalence with transport of electrons in a phonon medium}

In order to better clarify the role of the Boltzmann equation with
the inelastic collisional integral only,
we shall show now that this equation is mathematically equivalent to the transport
equation for electrons in a phonon background. Indeed, the results obtained in
the previous section are surprisingly similar with those found by Majorana in the
field of electrons transport in a semiconductor \cite{Majorana1,Majorana2}.\\
First of all, let us rewrite Eq. (\ref{Boltzmann}) with $\left(\partial\,f/\partial\,t\right)^{\rm el }_{\rm
coll}=0$, in the following form
\begin{equation}
{\partial\,\tilde f({\bfm w})\over\partial\,t}+\epsilon\,{\bfm
w}\cdot{\partial\,\tilde f({\bfm w})\over\partial\,{\bfm x}}+({\bfm
e}+{\bfm w}\times{\bfm b})\cdot{\partial\,\tilde f({\bfm
w})\over\partial\, {\bfm w}}=\left({\partial\,\tilde f\over\partial\,t}\right)^{\rm in}_{\rm
coll} \
,\label{tb}
\end{equation}
where we have posed $\bfm w={\bfm v}/\epsilon$, ${\bfm e}={\cal Q}\,{\bfm
E}/(m\,\epsilon)$, ${\bfm b}={\cal Q}\, {\bf B}/(m\,\epsilon)$ and
$\tilde f(\bfm w)=f({\bfm v})$.\\ The
collisional integral is now
\begin{eqnarray}
\nonumber
\left({\partial\,\tilde f\over\partial\,t}\right)^{\rm in}_{\rm
coll}&=&w^{-1}\int\Big[{\cal N}_1\,w_+^2\, \tilde I^{\rm
in}(w_+,\,{\bfm \Omega}\cdot{\bfm \Omega}^\prime)\,f(w_+\,
{\bfm\Omega^\prime})\\
\nonumber
& &+{\cal N}_2\,w^2\,\tilde I^{\rm in}(w,\,{\bfm
\Omega} \cdot{\bfm \Omega}^\prime)\,\tilde
f(w_-\,{\bfm\Omega^\prime})\,U(w_-^2)\Big]
\,d{\bfm\Omega^\prime}\\
\nonumber
& &-w^{-1}\,\tilde f({\bfm w})\int\Big[{\cal N}_2\,w_+^2\,\tilde
I^{\rm in}(w_+,\,{\bfm \Omega} \cdot{\bfm \Omega^\prime})\\
& &+{\cal
N}_1\,w^2\,\tilde I^{\rm in}(w,\,{\bfm \Omega}
\cdot{\bfm\Omega^\prime})\,U(w_-^2)\Big]\,d{\bfm \Omega^\prime} \
,\label{tiin}
\end{eqnarray}
where
$w_\pm=\sqrt{w^2\pm1}$ and $\tilde I^{\rm in}(w,\,{\bfm\Omega}\cdot{\bfm\Omega^\prime})=\epsilon\,I^{\rm in}(v,\,{\bfm\Omega}\cdot{\bfm\Omega^\prime})$.\\
By taking into account the relation
\begin{equation}
U(w\pm 1)\,\delta(w'-w_\pm)=2\,w_\pm\,\delta(w'^2-w^2\mp1) \ ,
\end{equation}
we may express $\left(\partial\,f/\partial\,t\right)^{\rm in}_{\rm
coll}$ in the following
equivalent form

\begin{eqnarray}
\nonumber
\left({\partial\,\tilde f\over\partial\,t}\right)^{\rm in}_{\rm
coll}&=&\int\Big[{\cal E}\,{\cal K}(w,\,w^\prime,\,{\bfm
\Omega} \cdot{\bfm\Omega^\prime})\,\delta({w^\prime}^2-w^2-1)\\
\nonumber
& &+{\cal
K}(w^\prime,\,w,\,{\bfm \Omega}
\cdot{\bfm\Omega^\prime})\,\delta(w^2-{w^\prime}^2-1)\Big]\,\tilde f({\bfm w^\prime})\,d{\bfm w^\prime}\\
\nonumber & &-\tilde f({\bfm w})\int\Big[{\cal
K}(w,\,w^\prime,\,{\bfm \Omega}
\cdot{\bfm\Omega^\prime})\,\delta({w^\prime}^2-w^2-1) \\
& &+{\cal E}\,{\cal K}(w^\prime,\,w,\,{\bfm \Omega}
\cdot{\bfm\Omega^\prime})\,\delta({w^\prime}^2-w^2-1)\Big]\,d{\bfm w^\prime} \ ,
\end{eqnarray}
where
\begin{equation}
{\cal K}(w,\,w^\prime,\,{\bfm \Omega}
\cdot{\bfm\Omega^\prime})=2\,{\cal N}_2\,{w^\prime\over w}\,\tilde
I^{\rm in}(w^\prime,\,{\bfm \Omega} \cdot{\bfm\Omega^\prime}) \ .
\end{equation}
In Ref. \cite{Majorana1}, free electrons have been considered
interacting with monochromatic phonons (energy $\hbar\omega$) of a semiconductor lattice.
The kinetic equation for electrons reported in that paper turns out
to have exactly the same form as Eq. (\ref{tb}), when the last
expression for $\left(\partial\,f/\partial\,t\right)^{\rm in}_{\rm
coll}$ is adopted, where $\Delta E$ plays the role of  $\hbar\omega$.

\sect{Conclusions}

Starting from the Boltzmann picture, we have studied a system of charged particles spreading in a
medium made by two-level atoms in the presence of an external electric and/or magnetic field.
For the two meaningful interaction laws of hard sphere and Maxwellian particles
we derived two wide families of distribution functions.\\
Concerning the hard sphere interactions, while the appearance of generalized Pearson
distribution is fascinating, we must summarize the assumptions we utilized to derive
this result: 1) the assumption that $M$ is much greater than $m$, which leads to the
linear equation for inelastic scattering; 2) the assumption that the electric field is small,
so that the P$_1$ approximation is allowed; 3) the assumption that $\Delta E$ is much smaller
than the thermal energy of a TP, so that the Fokker-Planck approximation is allowed;
4) the assumption that the cross sections are governed by power-laws.\\
On the other hand, concerning the Maxwellian interactions, we would like to remark that the statistical parameter $q$, for the power-law distribution in the Tsallis form, can actually be derived
strictly from microscopic arguments.
This circumstance, encountered in few models in which the Tsallis distribution
arises, is a very important point, related to the physical validity of this distribution, which is often argument of intensive debate \cite{Dauxois,Tsallis1}.

\app\sect{}

In this appendix we give some details about the distributions
derived in  the section V.\\

\noindent{\em Hard sphere interaction}\\
The hard sphere interaction follow by assuming constant values for
all cross sections, with $\sigma^{\rm el}(\xi)\equiv\sigma^{\rm
el}$ and $\sigma^{\rm in}(\xi)\equiv\sigma^{\rm in}$. In this case,
Eq. (\ref{distr}) takes the form of a generalized Pearson
distribution
\begin{equation}
F_1(\xi)=C_1\,\Big(1+b_1\,\xi+b_2\,\xi^2\Big)^{1/b_3}\,\exp\Big(-b_4\,\xi-b_5\,
\arctan(b_6+b_7\,\xi)\Big) \ ,
\end{equation}
whose coefficients $b_1,\,\ldots,\,b_7$, are related to
the microscopical parameters of the system according to
\begin{eqnarray}
\nonumber
b_1&=&{\beta_1\over\beta_2} \ ,\\
\nonumber
b_2&=&8\,{k\,T\over M}\,{\sigma^2\,\sigma^{\rm el}\over\beta_2} \ ,\\
\nonumber
b_3&=&12\,{(k\,T)^2\over m\,M}\,{\sigma\,\sigma^{\rm el}\over e^2} \ ,\\
b_4&=&{m\over2\,k\,T} \ ,\\ \nonumber
b_5&=&{m\over2\,k\,T}\,\left[{M\over3\,k\,T}\,{e^2\,\beta_1\over\sigma\,\sigma^{\rm
el}}-{16\over3}\,{({\bfm e}\cdot{\bfm
b})^2\over\sigma}\right]\,{1\over\beta_3} \ ,\\
\nonumber b_6&=&{\beta_1\over\beta_3} \
,\\
\nonumber b_7&=&16\,{k\,T\over M}\,{\sigma^2\,\sigma^{\rm
el}\over\beta_3} \ ,
\end{eqnarray}
with
\begin{eqnarray}
\nonumber \beta_1&=&{8\over3}\,e^2\,\sigma+8\,{k\,T\over
M}\,b^2\,\sigma^{\rm el}+\eta^2\,\sigma^2\,\sigma^{\rm in}_0 \
,\\
\beta_2&=&{8\over3}\,{({\bfm e}\cdot{\bfm
b})^2\over\sigma}+\eta^2\,b^2\,\sigma^{\rm in}_0 \ ,\\
\nonumber \beta_3&=&\left(32\,{k\,T\over
M}\,\sigma^2\,\sigma^{\rm el}\,\beta_2-\beta_1^2\right)^{1/2}
 \ .
\end{eqnarray}
The following particular cases arise:\\
1) For $\sigma^{\rm el}=0$, we obtain
the modified power-law distribution
\begin{equation}
F_2(\xi)=C_2\,\Big(1+c_1\,\xi\Big)^{-c_2}\,\exp\left(-c_3\,\xi\right)
\ ,\label{d2}
\end{equation}
with
\begin{eqnarray}
\nonumber c_1&=&{\sigma^2\over b^2}\,\left[1+{8\over3}\,{\big({\bfm
e}\cdot{\bfm
b}\big)^2\over \eta^2\,b^2\,\sigma\,\sigma_0^{\rm in}}\right]^{-1}\gamma \ ,\\
c_2&=&{4\over3}\,{m\over k\,T_\gamma}\,{e^2\,b^2-\big({\bfm e}\cdot{\bfm
b}\big)^2\over\gamma\,\eta^2\,\sigma^3\,\sigma_0^{\rm in}} \ ,\\
\nonumber
c_3&=&{m\over2\,k\,T_\gamma} \ ,
\end{eqnarray}
where
\begin{equation}
T_\gamma=\gamma\,T \ ,\quad\quad{\rm and}\quad\quad
\gamma=1+{8\,e^2\over3\,\eta^2\sigma\,\sigma_0^{\rm in}} \ .
\end{equation}
2) For $b=0$, we obtain the Margenau-Druyvesteyn distribution
\begin{equation}
F_3(\xi)=C_3\,\Big(1+d_1\,\xi\Big)^{d_2}\,\exp\left(-d_3\,\xi\right) \
,
\end{equation}
with
\begin{eqnarray}
\nonumber d_1&=&3\,{k\,T\over M}\,{\sigma\,\sigma^{\rm el}\over
e^2+{3\over8}\,\eta^2\sigma\,\sigma^{\rm in}_0} \ ,\\
d_2&=&{m\,M\over
6\,(k\,T)^2}\,{e^2\over\sigma\,\sigma^{\rm el}} \
,\\
\nonumber
d_3&=&{m\over2\,k\,T} \ ,
\end{eqnarray}
3) Again, for $b=0$ but without the elastic collisions
($\sigma^{\rm el}=0$), we obtain the
heated-Maxwellian distribution
\begin{equation}
F_4(\xi)=C_4\,\exp\left(-{m\,\xi\over2\,k\,T_\ast}\right) \ ,
\end{equation}
with
\begin{equation}
T_\ast=T\,\left(1+{8\,e^2\over3\,\eta^2\,\sigma\,\sigma^{\rm in}_0}\right)
\ .
\end{equation}

\noindent{\em Maxwellian interaction}\\
The Maxwellian interaction follow by assuming $\sigma^{\rm
el}(\xi)=\tilde\sigma^{\rm el}/\sqrt{\xi}$ and $\sigma^{\rm
in}(\xi)=\tilde\sigma^{\rm in}/\sqrt{\xi}$. The function (\ref{distr})
takes the form of a modified power-law distribution
\begin{equation}
F_5(\xi)=C_5\,\Big(1+g_1\,\xi\Big)^{-g_2}\,\exp\left(-g_3\,\xi\right) \ ,
\end{equation}
whose coefficients $g_1,\,g_2$ and $g_3$ are related to the
microscopical parameters of the system in
\begin{eqnarray}
\nonumber g_1&=&8\,{k\,T_\delta\over M}\,{\tilde\sigma^{\rm
el}\over\eta^2\,\tilde\sigma^{\rm in}_0} \ ,\\
g_2&=&{m\,M\over16\,(k\,T_\delta)^2}\,\delta\,\eta^2\,\tilde\sigma^{\rm el}\,\tilde\sigma^{\rm in}_0 \ ,\\
\nonumber
g_3&=&{m\over2\,k\,T_\delta} \ ,
\end{eqnarray}
where
\begin{equation}
T_\delta=(1+\delta)\,T \ ,\quad\quad{\rm and}\quad\quad
\delta={M\over3\,k\,T}\,{({\bfm e}\cdot{\bfm b})^2+e^2\,
\tilde\sigma^2\over (b^2+\tilde\sigma^2)\,\tilde\sigma\,\tilde\sigma^{\rm el}} \ .\label{td}
\end{equation}
The following particular cases arise:\\
1) For $\tilde\sigma^{\rm el}=0$ and $\tilde\sigma=\tilde\sigma^{\rm in}$, we obtain
the power-law distribution
\begin{equation}
F_6(\xi)=C_6\,\Big(1+h_1\,\xi\Big)^{-h_2} \ ,\label{ts}
\end{equation}
with
\begin{eqnarray}
h_1={8\over3}\,{({\bfm e}\cdot{\bfm b})^2+e^2\,
\tilde\sigma^2\over \eta^2\,(b^2+\tilde\sigma^2)\,\tilde\sigma\,\tilde\sigma^{\rm in}_0} \
,\quad\quad{\rm and}\quad\quad h_2={m\over2\,k\,T\,h_1} \ .
\end{eqnarray}
Distribution (\ref{ts}) can be written in the form (\ref{tsa}), by introducing
the deformed parameter $q=1+1/h_2$, given by
\begin{equation}
q=1+{16\over3}\,{k\,T\over m}\,{({\bfm e}\cdot{\bfm b})^2+e^2\,
\tilde\sigma^2\over \eta^2\,(b^2+\tilde\sigma^2)\,\tilde\sigma\,\tilde\sigma^{\rm in}_0} \ .
\end{equation}
2) For $\tilde\sigma^{\rm in}=0$ and $\tilde\sigma=\tilde\sigma^{\rm el}$,  we obtain
the heated Maxwellian distribution
\begin{equation}
F_7(\xi)=C_7\,\exp\left(-{m\,\xi\over2\,k\,T_\zeta}\right) \ ,
\end{equation}
where
\begin{equation}
T_\zeta=(1+\zeta)\,T \ ,\quad\quad{\rm and}\quad\quad\zeta={M\over3\,k\,T}\,{({\bfm e}\cdot{\bfm b})^2+e^2\,
\tilde\sigma^2\over (b^2+\tilde\sigma^2)\,\tilde\sigma^2} \ .
\end{equation}


\end{document}